# Reversing Ostwald Ripening

V. M. Burlakov[1], M. S. Bootharaju[2], T. M. D. Besong[2], O. M. Bakr[2], and A. Goriely[1]

[1]*Mathematical Institute, University of Oxford, Woodstock Road, Oxford, OX2 6GG, UK*
[2]*Division of Physical Sciences and Engineering, Solar and Photovoltaics Engineering Center, King Abdullah University of Science and Technology (KAUST), Thuwal 23955-6900, Saudi Arabia*

**Abstract**

The phenomenon of Ostwald Ripening is generally considered a limiting factor in the monodisperse production of nanoparticles. However, by analysing the free energy of a binary A-B solution with precipitated A-particles we show that there is a region in the parameter space of component concentrations and interaction energies where smaller particles are more stable than bigger ones. The strong binding of B-species to surfaces of A-particles significantly decreases the particles' effective surface energy, making it negative. The global minimum of free energy in such a system is thus reached when mass is transferred from bigger particles to the smaller ones, such that all particles become identical in size. The process of mass transfer is opposite to Ostwald ripening, and can be used for generating mono-disperse arrays of nanoparticles.

Microstructure coarsening, or Ostwald ripening (Or) [1], is a final stage of many first order phase transformations following nucleation and growth stages, and is often observed in two-phase mixtures [2], binary alloys [3], clusters on surfaces [4], oil-water emulsions [5], and during epitaxial growth [6-7] and synthesis of nanoparticles [8]. Ostwald ripening represents the spontaneous evolution of precipitated clusters, during which smaller clusters of atoms/molecules transfer mass to bigger clusters. The main driving force for Ostwald ripening is the minimization of the interfacial area between the two phases and as such depends intimately on the interface energetics and the system temperature.

Typically, the net effect of Ostwald ripening is to broaden the size distribution of nucleated particles [9-12]. As ripening takes place in liquid [13-16] and gas-phase [17-20] fabrication of nanoparticles, and epitaxial growth of quantum dot arrays [21-23], it prevents the highly desirable formation of narrow size distribution of particles. Indeed, identical semiconductor nanoparticles (quantum dots) would be extremely important for electro-optics, quantum information processing and many other applications where narrow spectrum excitation is required. The known mechanisms discussed in the literature for narrowing a particle size distribution (PSD) are: *i*) digestive ripening [13-15], and *ii*) inverse ripening [26-28]. Digestive ripening can be explained by taking into account electric charges on the particle and by including the electrostatic energy in the free energy minimum [24, 25]. Inverse ripening has been described in gold inclusions in an amorphous $SiO_2$ matrix irradiated with MeV gold ions [27, 28]. The narrowing of PSD by this mechanism is of transient character and requires fine tuning of radiation-annealing cycles.

In this Letter we present a new theoretical mechanism for narrowing a PSD by considering a simple dilute solution of two species (called *molecules*) in a neutral solvent in the presence of precipitated aggregates (called *particles*) of both components. Based on a free energy analysis, we show that there exists a range of parameters at which reverse coarsening (called here reverse Ostwald ripening, rOr) occurs, and consequently, smaller precipitated particles grow at the expense of the bigger ones. This regime is spontaneous, *i.e.* it is thermodynamically driven and converts *any* initial distribution of particles into an array of identical particles. The transition of the system to this regime can be driven by simply changing the temperature.

More explicitly, we consider a conservative system containing $M_A$ units of material A, $M_B$ units of material B and $K$ units of solvent. We refer to these units as molecules and particles with the understanding that the method is applicable to any system with the same characteristics. We define by $Q_{A(B)}$ any physical quantity $Q$ that is applicable to either material. For instance, phases A and B have precipitated spherical particles of radii $R_{A(B)}$, which contain of $P_{A(B)} = \dfrac{4\pi R_{A(B)}^3}{3a_{A(B)}^3}$ molecules, where $a_{A(B)}$ is a typical intermolecular distance between molecules. We assume that the A and B molecules bind to each other with binding energy $\varepsilon_{AB}$ in both the solution and on the precipitated particle surfaces. This binding in the solution results in the formation of *C* complexes A-B (dimers), while the binding on particle surfaces results in concentration $s_{BA}$ of B-molecules on A-particles and *vice versa*, a concentration $s_{AB}$ of A-molecules on B-particles. The Gibbs free energy of the system is

$$\begin{aligned}
G = &-\varepsilon_A P_A + \gamma_A S_A - \varepsilon_{AB} s_{BA} S_A + T \cdot \left[ s_{BA} S_A \ln(s_{BA}) + S_A \cdot (1 - s_{BA}) \ln(S_A \cdot (1 - s_{BA})) \right] \\
&-\varepsilon_B P_B + \gamma_B S_B - \varepsilon_{AB} s_{AB} S_B + T \cdot \left[ s_{AB} S_B \cdot \ln(s_{AB}) + S_B \cdot (1 - s_{AB}) \ln(S_B \cdot (1 - s_{AB})) \right] \\
&-\varepsilon_{AB} \cdot C + T \cdot \left( C \ln(C) + N_A \ln(N_A) + N_B \cdot \ln(N_B) + K \cdot \ln(K) - Z \ln(Z) \right),
\end{aligned} \quad (1)$$

where $\varepsilon_{A(B)}$ and $\gamma_{A(B)}$ is the cohesive and surface energies per A or B-molecule, respectively, $T$ is the temperature in energy units (eV), $S_{A(B)} = 4\pi R_{A(B)}^2 / a_{A(B)}^2$ is the number of molecules on the corresponding particle surface, $N_{A(B)} = M_{A(B)} - P_{A(B)} - S_{B(A)} s_{AB(BA)} - C$, $Z = N_A + N_B + C + K$. In Eq. (1), we assumed a simple form for the entropy, which provides a lower limit estimate [29]. The minimization of the free energy with respect to independent variables $P_A, P_B, C, s_{AB}$ and $s_{BA}$ in the limit of a dilute solution with $K \gg M_A + M_B$ yields

$$\begin{aligned}
T \ln(n_{A(B)}) &= -\varepsilon_{A(B)} + \frac{2a_{A(B)}}{R_{A(B)}} \cdot \Gamma_{A(B)}, \\
c &= E \cdot n_A \cdot n_B, \\
s_{AB} &= E \cdot n_A \cdot (1 + E \cdot n_A)^{-1}, \\
s_{BA} &= E \cdot n_B \cdot (1 + E \cdot n_B)^{-1},
\end{aligned} \quad (2)$$

where $E = \exp(\varepsilon_{AB}/T)$, and $\Gamma_{A(B)} = \gamma_{A(B)} - T \ln(1 + E \cdot n_{B(A)})$ is the effective surface energy of A(B)-particle, $n_{A(B)} = N_{A(B)}/Z$ and $c = C/Z$. The first two equations of (2) is a statement on the equality of chemical potentials for A(B)-molecules in the solution $\mu_{A(B)}(sol) = T \ln(n_{A(B)})$ and in the particle $\mu_{A(B)}(part) = -\varepsilon_{A(B)} + \frac{2a_{A(B)}}{R_{A(B)}} \cdot \Gamma_{A(B)}$. Note that the surface-energy term $\gamma_{A(B)}$ is renormalized by the entropic contribution from the B(A)-component and the strength of this renormalization depends on concentration of B(A)-molecules in the solution and on the A-B binding energy. The second and third lines in Eqs (2) give the equilibrium concentrations of A-B complexes in the solution and on the particle surfaces, respectively.

We are interested in the conditions leading to reverse Ostwald ripening (rOr) of A-particles, which in terms of molecular chemical potentials means that $\frac{d}{dR}\mu_A(part) > 0$. The latter is fulfilled if $\Gamma_A < 0$, that is

$$n_B > 1/E \cdot \left( \exp\left(\frac{\gamma_A}{T}\right) - 1 \right) \quad (3)$$

This inequality provides a condition on the concentration of B-molecules in solution so that rOr occurs for A-particles. In essence, the B-molecules act as surfactants passivating the A-particle's surfaces and sufficiently decreasing their effective surface energy so that it becomes negative. At negative surface energy the particles have a tendency of increasing the total surface-to-volume ratio, which is achieved by dissolving the bigger particles, whereas the smaller particles grow to reduce the chemical potential of A-molecules in the solution towards its equilibrium value. There could be even nucleation of new particles, as the nucleation barrier is decreased due to a decrease in the effective surface energy. The fact that the state with identical particles corresponds to the true equilibrium can be easily checked by calculating $\frac{d}{dR}\frac{\partial G}{\partial P_A}$, which turns out to be positive for any $R_A/a_A > 1$ if Eq. (3) is fulfilled.

We must add an extra to Eq. (3) condition for rOr by setting a lower limit for A-molecule concentration to ensure nucleation of A-particles. This requirement suggests that the chemical potential of A-molecules in the absence of nucleated A-particles $\mu_{A0} = T\ln(m_A - c_0)$ must be higher than $\mu_A(R)$, i.e.

$$T\ln(m_A - c_0) > -\varepsilon_A + \frac{2a_A}{R_A}\cdot\Gamma_A, \tag{4}$$

where $m_A = M_A/Z$ and $c_0$ is the concentration of A-B complexes in the absence of particle, *i.e.* when $P_{A(B)} = s_{AB(BA)} = 0$. Since $\Gamma_A \approx 0$ at the border between the normal ripening and the rOr, we have

$$\ln(m_A - c_0) > -\varepsilon_A/T. \tag{5}$$

The concentration $c_0$ is given by Eq. (2) as $c_0 = E\cdot n_{A0}\cdot n_{B0} = E\cdot(m_A - c_0)\cdot(m_B - c_0)$. In the dilute approximation that we have assumed ($c_0 \ll 1$), the solution to this equation is

$$c_0 \approx m_A m_B/(m_A + m_B + 1/E). \tag{6}$$

Substituting this into Eq. (5) and using Eq. (3) we obtain the condition

$$m_A > -\frac{1}{2}\left[1-\exp\left(\frac{\varepsilon_{AB}-\varepsilon_A}{T}\right)\right] + \frac{1}{2}\sqrt{\left[1-\exp\left(\frac{\varepsilon_{AB}-\varepsilon_A}{T}\right)\right]^2 + 4\exp\left(\frac{\gamma_A - \varepsilon_A}{T}\right)} \tag{7}$$

The last condition for rOr can be interpreted as a condition on the maximum concentration of A-molecules. These molecules act as surfactant and therefore should not precipitate, i.e. their concentration should be below solubility limit. This is achieved by imposing a similar but opposite condition opposite to that given by Eq. (5) for A-particles: $\ln(m_B - c) < -\varepsilon_B/T$. If the total numbers of A and B molecules are roughly the same and a significant fraction of A-molecules is precipitated without precipitation of B-molecules then $c \ll m_B$ and $n_B = m_B - c \approx m_B$, or $m_B < \exp(-\varepsilon_B/T)$. Therefore, the conditions to have rOr in the system of particles (A) in the presence of surfactant molecules (B) can be written as

$$\exp(-\varepsilon_B/T) > m_B > \exp\left(\frac{\gamma_A - \varepsilon_{AB}}{T}\right) - \exp\left(-\frac{\varepsilon_{AB}}{T}\right),$$

$$m_A > -\frac{1}{2}\left[1 - \exp\left(\frac{\varepsilon_{AB} - \varepsilon_A}{T}\right)\right] + \frac{1}{2}\sqrt{\left[1 - \exp\left(\frac{\varepsilon_{AB} - \varepsilon_A}{T}\right)\right]^2 + 4\exp\left(\frac{\gamma_A - \varepsilon_A}{T}\right)} \tag{8}$$

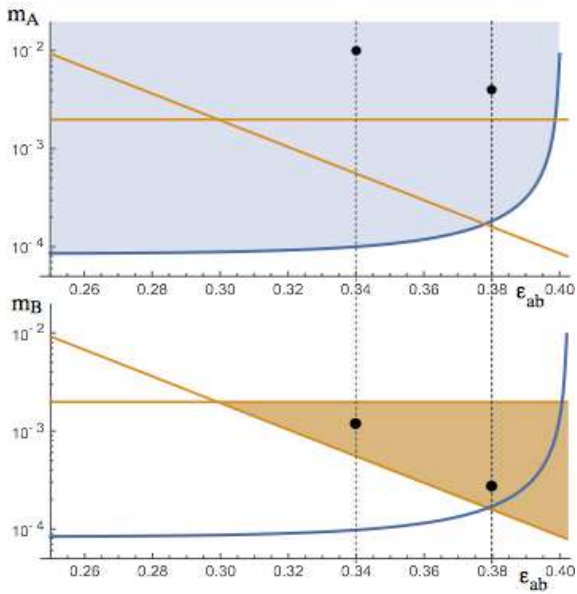

Fig. 1. Regions in parameter space ($\varepsilon_{AB}$ and $m_A = m_B$) where reverse Ostwald ripening (rOr) occurs, according to the conditions (8) for T=0.032 eV (100° C), $\gamma_A$=0.1 eV, $\varepsilon_A$=0.4 eV, $\varepsilon_B$=0.2 eV. The shaded region in top graph represents the condition on $m_A$, while the bottom graph is the condition on $m_B$. For a given value of $\varepsilon_{AB}$ (say 0.34 or 0.38), different values of $m_A$ and $m_B$ (indicated by symbols) falling into corresponding shaded areas lead to rOr.

The conditions for rOr are illustrated in Fig. 1 for two temperatures T=50 C (a) and T=100 C (b) by the doubly shaded areas in the parameter space $\varepsilon_{AB}$ and $m_A$, $m_B$.

So far the theory was developed with the assumption that all particles are spherical and their surface energy does not depend on their radius. Typically, particles' (or clusters') are not spherical and the deviation from sphericity is size-dependent. To take this effect into account we consider the energy (per atom) $W$ of a small cluster containing $N$ atoms. The dependence of this energy upon $N$ can be described (at least for metallic clusters) as [30-32]

$$W = A + B \cdot N^{-1/3} + C \cdot N^{-2/3} + D \cdot N^{-1} \tag{9}$$

where $A<0$ is the bulk contribution to energy (cohesive energy), $B>0$ is the surface contribution, $C$ is the contribution of edges, and $D$ defines the energy origin (reference point). Adopting this expression for the generic case of sphere-like cluster of A-atoms with typical radius $R = R_A / a_A$ we use $N \sim R^3$ in (9) so that

$$-\varepsilon_A(R) \approx -\varepsilon_A + \frac{2}{R} \cdot \gamma_A(R) \qquad (10)$$

where $\gamma_A(R) = \gamma_{A0} + \frac{\gamma_{A1}}{R}$, and we have neglected small energy origin $D$ [33]. The deviation of a cluster's shape from spherical can be taken into account by the size dependence of its surface energy. Both parameters $\gamma_{A0}$ and $\gamma_{A1}$ are positive, as the former corresponds to true surface energy for $R \gg 1$, while the latter takes into account the relative increase of surface area for non-spherical clusters reflecting the fact that this increase is higher for smaller clusters. Therefore, the surface energy $\gamma_A(R)$ decreases with increasing cluster radius making larger clusters relatively more stable. The effective surface energy in the rOr regime is negative, and the increasing $R$ implies that bigger clusters are less stable (see (2) and the text below it). This means that for any $\varepsilon_A$, $\varepsilon_B$, $\gamma_{A0}$ and $\gamma_{A1}$ there is an optimum value of $R$ corresponding to the minimum of atomic chemical potential [34]

$$\mu_A(R) = -\varepsilon_A + \frac{2}{R} \cdot \left( \gamma_{A0} + \frac{\gamma_{A1}}{R} - T \ln(1 + E \cdot n_B) \right), \qquad (11)$$

which gives the cluster radius at equilibrium

$$R_{eq} = -\frac{2\gamma_{A1}}{\gamma_{A0} - T \ln(1 + E \cdot n_B)} \qquad (12)$$

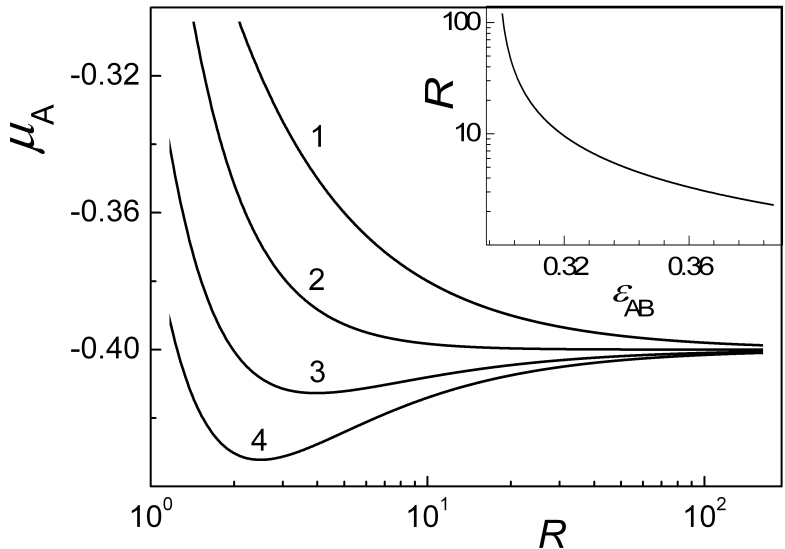

Fig. 2. Atomic chemical potential $\mu_A$ as a function of cluster radius $R$ at T=0.032 eV (100° C), $\gamma_{A0} = \gamma_{A1} = 0.1$ eV, $\varepsilon_A = 0.4$ eV, $\varepsilon_B = 0.2$ eV and $n_B \approx 0.002$ (solubility limit for B) for various values of $\varepsilon_{AB}$ (in eV): 1 - $\varepsilon_{AB} = -0.3$ (no surface passivation); 2 - $\varepsilon_{AB} = 0.3$; 3 - $\varepsilon_{AB} = 0.34$; 4 - $\varepsilon_{AB} = 0.38$. Insert shows $R_{eq}$ as a function of $\varepsilon_{AB}$. rOr corresponds to situation when $\mu_A$ has minimum at finite $R$ values (curves 3 and 4).

The functions $\mu_A(R)$ and $R_{eq}(\varepsilon_{AB})$ are presented in Fig. 2, which shows that a minimum in $\mu_A(R)$ is observed only for $\varepsilon_{AB} > \varepsilon_{Crit}$, where $\varepsilon_{Crit}$ is determined by setting the denominator in Eq. (12) to zero and found to be ~0.3 eV. The minimum in $\mu_A(R)$ is well pronounced only at $\varepsilon_{AB} \geq 0.34$ eV (curves 3 and 4), the values approaching $\varepsilon_A$. Note, that $\varepsilon_{AB} \leq \varepsilon_A$ as otherwise no nucleation of A-clusters would take place.

As can be seen from Fig. 2, the driving force for the system towards an energy minimum is rather strong for small clusters with $R<R_{eq}$ and significantly weaker for larger clusters with $R>R_{eq}$ especially at relatively lower values of $\varepsilon_{AB}$. Furthermore, smaller clusters are in a regime of ordinary ripening while the bigger ones may enter the rOr regime. Therefore it is worth checking the overall effect of coarsening by direct simulation of the process. We assume that there is a distribution of clusters nucleated and grown so that further evolution is only possible by exchange of A-atoms between clusters. The cluster surfaces are covered with B-molecules, which significantly slows down an exchange of A-atoms between the clusters and solution. Therefore it is reasonable to assume that the atomic diffusion in solution is the fastest process and the ripening process is limited by the attachment-detachment events with effective reaction rate constant $K$. Then the evolution of $i$-th cluster with radius $R_i$ (measured in the units of inter-atomic distance) within this approximation (Wagner approximation [35]) can be described using the equation

$$\frac{dR_i}{dt} = K \cdot (n - n_{GT}(R_i)) \qquad (13)$$

Where $n_{GT}(R_i)$ is the Gibbs-Thomson concentration and $n$ is the mean field concentration of A-atoms in solution. The latter is determined using mass conservation for all clusters

$$\frac{d}{dt}\sum_{i=1}^{N}\frac{4}{3}\pi R_i^3 = \sum_{i=1}^{N} 4\pi R_i^2 \frac{dR_i}{dt} = \sum_{i=1}^{N} 4\pi K R_i^2 \cdot (n - n_{GT}(R_i)) = 0 \rightarrow n = \frac{\sum_{i=1}^{N} R_i^2 \cdot n_{GT}(R_i)}{\sum_{i=1}^{N} R_i^2} \qquad (14)$$

While the Gibbs-Thomson concentration is the concentration where the cluster is in equilibrium with the surrounding solution, i.e. it is determined from the first equation of (2) taking into account Eq. (11)

$$n_{GT}(R_i) = \exp\left[-\frac{\varepsilon_A}{T} + \frac{2\gamma_{A0}}{T \cdot R_i} + \frac{2\gamma_{A1}}{T \cdot R_i^2} - \frac{2}{R_i}\ln(1 + E \cdot n_B)\right] \qquad (15)$$

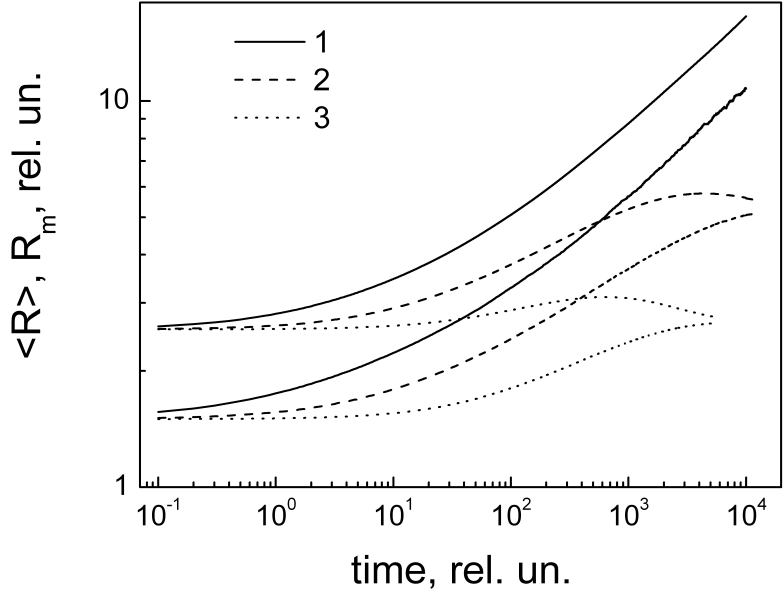

Fig. 3. Time evolution of the mean $\langle R \rangle$ and maximum $R_m > \langle R \rangle$ radii calculated for ensemble of $10^5$ particles with Gaussian initial distribution ($\langle R \rangle = 1.5$ and standard deviation $\sigma = 0.25$) obtained via numerical solution of Eqs (13) - (15) for T=0.032 eV (100° C), $\gamma_{A0} = \gamma_{A1} = 0.1$ eV, $\varepsilon_A = 0.4$ eV, $\varepsilon_B = 0.2$ eV and $n_B \approx 0.002$. The values of $\varepsilon_{AB}$ are: 1) 0.3 eV, 2) 0.34 eV, and 3) 0.38 eV.

Using Eqs (13)-(15) we simulated evolution of $10^5$ clusters for different values of $\varepsilon_{AB}$. The main results are summarized in Fig. 3. As one can see both the mean <R> and maximum $R_m$ radii continuously grow with time for $\varepsilon_{AB} = 0.30$, which is characteristic for ordinary Or. In contrast, for $\varepsilon_{AB} = 0.34$ eV and $\varepsilon_{AB} = 0.38$ eV the evolution is indicative of rOr: the radii initially grow and then <R> levels up while $R_m$ starts to decrease such that both converge to a single value. Note that the asymptotic values of the mean radii are close to those corresponding to the chemical potential minima in Fig. 2. It is also worth pointing out that a small variation in the initial distribution of particles does not affect final results. A significant change in the initial distribution may however affect the particle evolution as it depends on relative fraction of smaller particles which usually are in the regime of ordinary Or and their quick dissolution pushes bigger particles towards equilibrium radius (see Fig. 2) and even further. This may explain initial growth of $R_m$ over $R_{eq}$.

To illustrate the theoretical predictions we carried out synthesis of gold nanoclusters using glutathione as the surfactant and analysing their size distribution after 40 min, 2.5 hours, and 24 hours from the start of the synthesis (see Supporting Material for details). It was found that after 40 minutes the particles are polydisperse with mean diameter 2.89 nm and the variance ~ 2.0 nm (Fig. 1a). After 2.5 h mean particle diameter remains the same (2.88 nm) while the variance decreased more than twice – Fig. 1b. Even more monodisperse particles with mean diameter of 2.60 nm and the variance 0.3 nm were observed 24 h later (Fig. 1c). As illustrated in Fig. 1 the larger particles gradually decrease in size and the smaller ones disappear such that the entire distribution narrows with time. This behaviour of particle radii is very similar to that discussed in the previous paragraph and is indicative of the rOr effect.

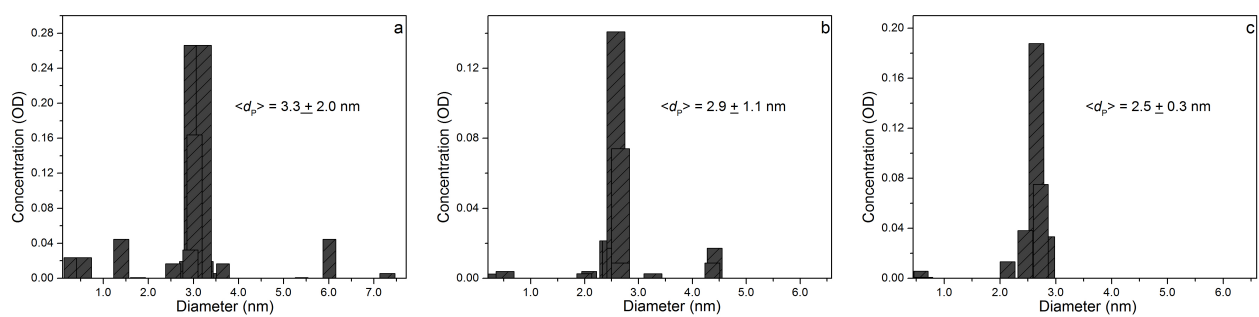

Fig. 4. Analysis of the particle size distribution in the synthesis of Au nanoclusters in aqueous solution by means of Sedimentation Velocity *via* analytical ultracentrifugation, AUC (see Supporting Material). The plots show the particle diameter distributions after (a) 40 min, (b) 2.5 h and (c) 24 h from the start of the synthesis.

In summary, we presented both free energy analysis and simulation studies of dilute binary molecular solution with precipitated particles of one component and the other component acting as surfactant passivating the particle surfaces. The analysis predicts that at certain component concentrations and strength of inter-molecular interaction the system of precipitated particles evolves towards equilibrium in such a way that the PSD spontaneously narrows resulting in eventually mono-sized array of particles. The reason for the narrowing of the size distribution is a strong enough surface passivation of particle surfaces due to bonding between the molecular components, which makes effective surface energy of particles negative. The effect is confirmed by direct simulation of ripening process in the large ensemble of particles and by the experimental resizing of small gold nanoclusters. The rOr effect can be used for solution-processed generation of mono-sized nanoparticulate arrays of various materials. The effect can also be used for surface treatment of bulk materials to produce highly developed roughness. It is worth noting that the presented analysis may explain digestive ripening without invoking any electrical charging of particles.

**Acknowledgements**

VB acknowledges support from the Oxford Martin Fellowship.

# Supporting Material

## Reversing Ostwald Ripening

V. M. Burlakov[1], M. S. Bootharaju[2], T. M. D. Besong[2], O. M. Bakr[2], and A. Goriely[1]

[1]*Mathematical Institute, University of Oxford, Woodstock Road, Oxford, OX2 6GG, UK*
[2]*Division of Physical Sciences and Engineering, Solar and Photovoltaics Engineering Center, King Abdullah University of Science and Technology (KAUST), Thuwal 23955-6900, Saudi Arabia*

**Experimental section**

**Chemicals**

All chemicals are commercially available and were used as purchased. Hydrogen tetrachloroaurate(III) trihydrate ($HAuCl_4 \cdot 3H_2O$, 99.99% from Alfa Aesar) and reduced L-glutathione (HSG, 97% from Sigma) were used as gold and thiol ligand precursors, respectively. Carbon monoxide was supplied by Specialty Gas Center (SGC), Jeddah, Saudi Arabia. Sodium hydroxide pellets were used as received.

**Synthesis of gold nanoclusters protected with glutathione (HSG)**

The synthesis of glutathione (HSG)-protected Au nanoclusters was carried out using the carbon monoxide (CO) reduction method [1]. Typically, about 23 mg of HSG was added into a round bottom flask containing 25 mL of 2 mM aqueous $HAuCl_4 \cdot 3H_2O$. The light yellow color of gold chloride solution turned colorless within 30 minutes of stirring at room temperature. A colorless solution indicates the formation of gold-thiolate complexes. The solution pH was adjusted to 11 using 1 M NaOH. Reduction was carried out with CO gas at a pressure of 30 psi for 3-4 minutes. The colorless reaction solution then changed to light yellow and finally to deep brown (after 4 hours) indicating the formation of nanoclusters. To ensure complete reduction, CO gas was bubbled through the solution for 24 hours. At specific time intervals, the solutions of clusters from various batches of synthesis (maintaining the same experimental conditions) were withdrawn using a syringe for sedimentation velocity in analytical ultracentrifugation (SV-AUC) analysis.

**Analytical Methods**

The sedimentation coefficient is often a useful parameter in the characterization of nanoparticles in various solvents as it is possible to accurately assess the evolution of the distribution of molecular species (or particles) with respect to time, shape (or molecular dimensions) and molecular weight. Sedimentation coefficients can be obtained through sedimentation velocity experiments during which the components of a given solution are separated as the molecular species of the solution sediment in the AUC cell. The process of sedimentation is monitored by measuring the absorption spectra of the solution over the entire length of the cell in the course of the experiment. The resulting data can be analyzed with a variety of methods (e.g., Sedfit, Ultrascan, Sedanal etc.) to obtain the sedimentation coefficient, from which the molecular dimensions of the various components can be derived. As even monodispersed nanoparticles are known to display a distribution of sedimentation coefficients owing to their inherent anisotropy[2, 3], we assumed an average particle frictional ratio of 1.45 for the different nanoparticle preparations in order to carry out a 2-dimensional spectrum analysis (2DSA) fit to the sedimentation velocity data. This results in a distribution of particle partial specific volume from which the particle density can be computed. The average particle diameter ($d_P$) of the distribution is then calculated from equation 1 below.

$$d_P = \sqrt{\frac{18\eta_s s}{\rho_P - \rho_s}} \quad (1)$$

Where $\eta$ is the solvent viscosity, $s$ is the sedimentation coefficient of the particle, $\rho_P$ is the particle density, and $\rho_s$ is the solvent density.

Since AUC achieves particle separation during high speed sedimentation, the molecular weight of the major sedimenting particle can also be obtained from the sedimentation velocity data by combining the particle and solution densities, the diffusion ($D$) and sedimentation ($s$) coefficients of the particle via the Svedberg equation (equation 2) [4].

$$M = \frac{sRT}{D}\left(1 - \frac{\rho_s}{\rho_P}\right)^{-1} \quad (2)$$

Where, $R$ is the gas constant and $T$ is the temperature in Kelvin.

**Instrumentation**

AUC experiments of all samples was performed using a Beckman Coulter XL-A analytical ultracentrifuge equipped with absorption optics. Sample and reference solutions (400 µL) were filled into a standard cell (12 mm path length) with titanium centerpieces and sapphire windows and subsequently inserted into a 4-hole An60Ti rotor (Beckman Coulter). All samples were run at 45000 rpm in order to collect as much information from the sedimenting particles as possible and scans were acquired in intensity mode at 450 nm. At least 60 scans were acquired for each sample and data analysis was performed with Ultrascan 3.3 (Revision 1884) using the 2-dimensional spectrum analysis model with 75 Monte Carlo iterations.

**Results and Discussion**

To understand the growth of the $Au_{25}(SG)_{18}$ clusters in solution clearly, we performed AUC measurements in an attempt to follow the distribution of molecular components in the nanoparticle solution within a 24 hour period. Sample nanocluster solutions at different stages (or times – 40 minutes, 2.5 hours and 24 hours) of synthesis were retrieved from the reaction flask and subjected to sedimentation velocity in the analytical ultracentrifuge (SV-AUC). A summary of the SV-AUC results is shown in Figure S1, which displays the evolution of the distribution of sedimentation coefficients ($s$) of the nanocluster particles during synthesis. It is observed that while the sedimentation coefficient of the major sedimenting particle is more or less similar over a 24 hour period, there are significant changes in the distribution of the solution particles within the same period. The sedimentation velocity analyses show a narrowing of the sedimentation coefficient distribution over time with the broadest distribution (~1 – 7 x $10^{-13}$ s) observed 40 minutes after the commencement of synthesis (Figure S1a) followed by a narrow distribution (~1 – 5 x $10^{-13}$ s) after 2.5 hours (Figure S1b) and a narrower distribution (~1 – 3 x $10^{-13}$ s) after 24 hours (Figure S1c). Tracking the average sedimentation coefficient of the entire distribution over time shows that the early stages of synthesis is characterized by a solution that is predominantly polydisperse after 40 minutes with an average particle sedimentation coefficient of 3.2 (±1.0) x $10^{-13}$ s. The average sedimentation coefficient remains fairly constant for the next 3 hours (3.0 (±1.0) x $10^{-13}$ s after 2.5 hours), but the range of sedimentation coefficients narrows suggesting that the solution is becoming less polydisperse with time. After 24 hours, the average sedimentation coefficient becomes more precise (3.0 (±0.3) x $10^{-13}$ s) indicating that the later stages of synthesis is dominated by the presence of nanoparticles with similar molecular conformations – a more monodisperse solution. The appearance of larger (with higher $s$ values) and smaller (with lesser $s$ values) molecular components in the initial stages and their gradual disappearance with time suggests a bidirectional growth mechanism, in which smaller intermediates grow into the final product ($Au_{25}SG_{18}$) [1] and highly unstable larger complexes unfold or disintegrate into the same final product [5]. The fact that these two sets of intermediates are underrepresented in the final solution compared to the beginning points to their defining roles during synthesis (see main text). The final sedimentation coefficient of the major sedimenting nanoparticle after 24 hours is 3.24 x $10^{-13}$ s (or 3.24 S) corresponding to an average molecular weight of 10700 Da.          Transformation of the sedimentation coefficient distribution into a particle size distribution using equation 1 above displays a similar trend as the distribution of sedimentation coefficients indicating the presence of a mix population of solution particles at the start of synthesis, which then narrows down into a less polydisperse solution after 24 hours (Figure S1 a-c top axes). For example, analysis of the solution 40 minutes after synthesis revealed particles with a mean average diameter of 3.3 ± 2.0 nm in a range of up to 7.3 nm (Figure S1a). This indicates that the nanoparticle solution was initially polydisperse having a mixture of small, intermediate and large particles. After 2.5 hours, the mean average particle diameter was similar (2.9 ± 1.1 nm), but the range of molecular diameters became less (up to 5.7 nm – Figure S1b) suggesting that as synthesis or growth of individual nanoparticles proceeds, the solution is becoming less polydisperse. A more monodisperse solution was obtained 24 hours later with a mean particle diameter of 2.5 ± 0.3 nm in a range of up to 3.5 nm (Figure S1c).

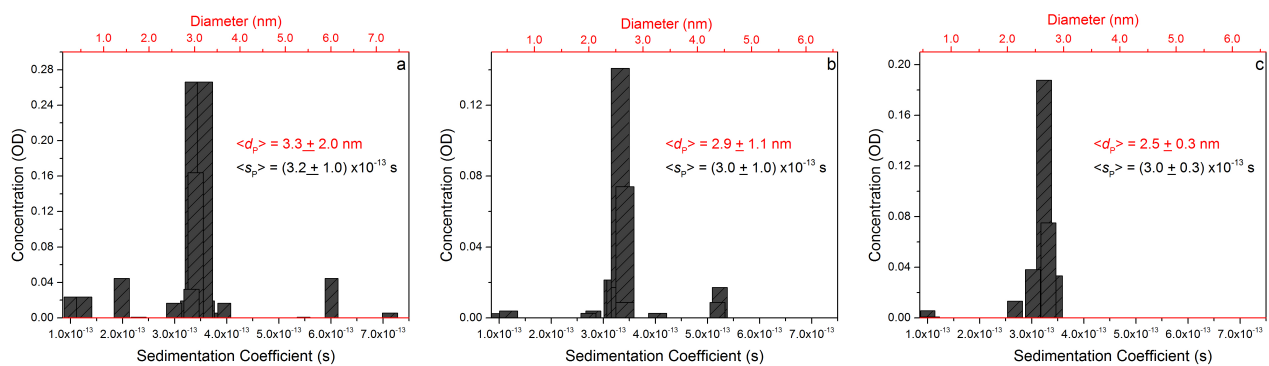

**Figure S1.** Analysis of molecular species distribution in the synthesis of $Au_{25}(SG)_{18}$ nanoclusters in aqueous solution by means of Sedimentation Velocity (SV) via analytical ultracentrifugation (AUC). The sedimentation coefficient distribution and corresponding molecular diameter distribution obtained through transformation of the sedimentation coefficient distributions using equation 1 is shown in the bottom and top axes, respectively, of each plot. The average sedimentation coefficient ($<s_P>$) and average molecular diameter ($<d_P>$) over the entire distribution is shown with the standard deviation reported as the standard error. The synthesis was followed after (a) 40 minutes, (b) 2.5 hours and (c) 24 hours.